\documentclass[conference]{IEEEtran}
\IEEEoverridecommandlockouts
\usepackage{cite}
\usepackage{amsmath,amssymb,amsfonts}
\usepackage{algorithmic}
\usepackage{bm}
\usepackage{graphicx}
\usepackage{textcomp}
\usepackage{xspace}
\usepackage{xcolor}
\usepackage{float}
\usepackage{cite}
\usepackage[final]{microtype}
\usepackage{mathtools}
\mathtoolsset{showonlyrefs=true}

\def\BibTeX{{\rm B\kern-.05em{\sc i\kern-.025em b}\kern-.08em
    T\kern-.1667em\lower.7ex\hbox{E}\kern-.125emX}}

\newtheorem{lemma}{Lemma}

\newtheorem{remark}{Remark}

\begin{document}

\title{On the Efficient Implementation of an Implicit Discrete-Time Differentiator\\
%\thanks{The authors sincerely thank CONACyT for the scholarship provided during this investigation to the student with No. CVU 555845, and to CINVESTAV for the provided resources.} 
}

% \author{
% \IEEEauthorblockN{
% J.~E. Carvajal-Rubio}
% \IEEEauthorblockA{\textit{Dept. of Electrical Engineering} \\
% \textit{and LAMIH, CNRS UMR 8201} \\
% \textit{CINVESTAV-IPN, Guadalajara}\\
% Zapopan, M\'exico. \\
% \textit{Polytechnic University of Hauts-de-France}\\
% Valenciennes, France. \\
% Jose.Carvajal@cinvestav.mx}

% \and

% \IEEEauthorblockN{J.~D. S\'anchez-Torres}
% \IEEEauthorblockA{\textit{Dept. of Mathematics and Physics} \\
% \textit{ITESO}\\
% Tlaquepaque, M\'exico \\
% dsanchez@iteso.mx}

% \and

% \IEEEauthorblockN{M. Defoort}
% \IEEEauthorblockA{\textit{LAMIH, CNRS UMR 8201} \\
% \textit{Polytechnic University of Hauts-de-France}\\
% Valenciennes, France \\
% michael.defoort@uphf.fr}

% \and

% \IEEEauthorblockN{A.~G. Loukianov}
% \IEEEauthorblockA{\textit{Dept. of Electrical Engineering} \\
% \textit{CINVESTAV-IPN, Guadalajara}\\
% Zapopan, M\'exico. \\
% alexander.loukianov@cinvestav.mx}

% \and

% \IEEEauthorblockN{M. Djemai}
% \IEEEauthorblockA{\textit{LAMIH, CNRS UMR 8201} \\
% \textit{Polytechnic University of Hauts-de-France}\\
% Valenciennes, France \\
% Mohamed.Djemai@uphf.fr}

% }

\author{\IEEEauthorblockN{J.~E. Carvajal-Rubio\IEEEauthorrefmark{1}\IEEEauthorrefmark{3},
J.~D. S\'anchez-Torres\IEEEauthorrefmark{2},
M. Defoort\IEEEauthorrefmark{3},
M. Djemai\IEEEauthorrefmark{3} 
and
A.~G. Loukianov\IEEEauthorrefmark{1}}
\IEEEauthorblockA{\IEEEauthorrefmark{1}Dept. of Electrical Engineering, CINVESTAV-IPN Guadalajara, Zapopan, M\'exico. \\
Email: Jose.Carvajal@cinvestav.mx, alexander.loukianov@cinvestav.mx}
\IEEEauthorblockA{\IEEEauthorrefmark{2}Dept. of Mathematics and Physics, ITESO, Tlaquepaque, M\'exico. \\
Email: dsanchez@iteso.mx}
\IEEEauthorblockA{\IEEEauthorrefmark{3}LAMIH, CNRS UMR 8201, Polytechnic University of Hauts-de-France, Valenciennes, France. \\
Email: michael.defoort@uphf.fr, Mohamed.Djemai@uphf.fr}
}

\pretolerance=8000
\tolerance=8000
\maketitle

\begin{abstract}
New methodologies are designed to reduce the time complexity of an implicit discrete-time differentiator and the simulation time to implement it. They rely on Horner's method and the Shaw-Traub algorithm. The algorithms are compared for differentiators of order 3, 7, and 10. The Half-Horner and Full-Horner methods showed the best performance and time complexity. 
\end{abstract}


\section{Introduction}

An online differentiator is useful for several applications, such as control laws based on derivatives of a signal, estimation of unmeasured states and parameters \cite{kaveh2008blood,shtessel2007smooth,iqbal2010robust}. The well-known homogeneous differentiator \cite{Levant_HSMD} was proposed in continuous-time and allow to estimate the first $n$ derivatives of a signal, if its $n$-th derivative has a known Lipschitz constant $L>0$. As the standard differentiator is implemented in digital systems \cite{Utkin2009}, a discrete-time version is implemented, for instance the explicit and implicit discrete-time realizations in \cite{Miki2014,Stefan_dif,koch2019discrete,barbot2020discrete,carvajal2021implicit}. Particularly, the implicit discrete-time differentiator, proposed in \cite{carvajal2021implicit}, has a remarkable reduction of the numerical chattering and preserves  the main properties of the continuous-time differentiator \cite{Levant_HSMD}, i.e., homogeneity property, asymptotic accuracy and convergence of its errors in finite-time to a vicinity of the origin. The main drawback of the implicit-discrete time differentiator is that the root of a polynomial has to be calculated almost each iteration for its implementation. The polynomial can not be calculated previous to its implementation because one of its parameters is updated each iteration. To implement this differentiator, in \cite{carvajal2021implicit} was proposed the Halley's method \cite{mcnamee2013numerical}, which converges to the unique positive root of the polynomial with $3$-order for a non-restrictive set of initial conditions.

Although Halley's method reduces the time required to implement the implicit differentiator, Halley's method needs the evaluation of the function and its first two derivatives. Then Halley's algorithm has a quadratic time with respect to $n$. Several algorithm has been proposed to reduce the number of basic operations (time complexity \cite{sipser2012introduction}) required to evaluate a polynomial and its derivatives, for instance, Horner's method \cite{mcnamee2007numerical}, Shaw-Traub algorithm \cite{shaw1974number}, and the De Jong Van algorithm \cite{de1975improved}. Other algorithms adapt or precondition parameters \cite{knuth2014art}, they will not be considered because in this work the polynomials are updated each iteration. On the other hand, the evaluation of the implicit differentiator, after calculate the respective roots, presents a cubic time, but it can be reduced to a quadratic time or linearithmic time using Horner's and the discrete Fourier transform \cite{golub2013matrix}, respectively.

The first contribution of this paper is to introduce new methodologies designed to implement the implicit discrete-time differentiator \cite{carvajal2019discretization} (HIDD), which allow to reduce the time complexity of the differentiator. The second one is a numeric comparative between them. This work is organized as follows. In Section \ref{sec:2Implicitdiscretetime}, implicit differentiator is defined. Additionally, Section \ref{sec:3Mainresults} contains the proposed algorithms for the implementation of implicit differentiator and its time complexity is calculated. Section \ref{sec:4Simulations} aims to compare the time complexity of the algorithms and the simulation time for its implementation. In Section \ref{sec:5Conclusion}, the main results of the paper are summarized and the future work is presented.

\section{Implicit Discrete-time Differentiator}\label{sec:2Implicitdiscretetime}
Let $\tau$ be the sampling time constant and defined as $\tau=t_{k+1}-t_k$ for $k=0,1,2,\cdots$. Then HIDD is defined by the following equations:

\begin{align}
    \begin{split}\label{eq:HIDD}
        &\bm{z}_{k+1}=\bm{\Phi} \left( \tau \right)\bm{z}_k+\bm{B}^\ast \left( \tau \right) \bm{v}\left( \widetilde{\sigma}_{0,k+1} \right),\\
        &\bm{v}\left( \widetilde{\sigma}_{0,k+1} \right)=\left[\Psi_{0,n}\left(\widetilde{\sigma}_{0,k+1} \right)  \; \;\cdots \; \; \Psi_{n,n}\left(\widetilde{\sigma}_{0,k+1} \right)  \right]^T, \\
        &\Psi_{i,n}(\widetilde{\sigma}_{0,k+1})=-\lambda_{n-i}L^{\frac{i+1}{n+1}} \left|\widetilde{\sigma}_{0,k+1}\right|^{\frac{n-i}{n+1}}\xi_{k}.
    \end{split}
\end{align}
For HIDD, $z_{i,k+1}$ corresponds to the estimation of $f^{i}_{0,k}$. On the other hand, $\bm{\Phi}\left( \tau \right)$ and $\bm{B}^\ast \left( \tau \right)$ have the following representation:
\begin{align} \label{eq:Phi}
    \begin{split}
        \bm{\Phi}\left( \tau \right)=\left[ \begin{array}{cccccc}
         1 & \tau & \frac{\tau^2}{2!} & \cdots & \frac{\tau^{n-1}}{\left( n-1 \right)!} & \frac{\tau^{n}}{n!} \\
         0 & 1 & \tau & \cdots & \frac{\tau^{n-2}}{\left( n-2 \right)!} & \frac{\tau^{n-1}}{\left( n-1 \right)!} \\
         \vdots & \vdots & \vdots & \ddots & \vdots & \vdots \\
         0 & 0 & 0 & \cdots & 1 & \tau \\
         0 & 0 & 0 & \cdots & 0 & 1 
         \end{array} \right],
    \end{split}
\end{align}
\begin{align}\label{eq:B}
    \begin{split}
        \bm{B^\ast}\left( \tau \right)=\left[ \begin{array}{cccccc}
        \tau & \frac{\tau^2}{2!} & \frac{\tau^3}{3!} & \cdots & \frac{\tau^{n}}{n!} & \frac{\tau^{n+1}}{\left( n+1 \right)!} \\
         0 & \tau & \frac{\tau^2}{2!} & \cdots & \frac{\tau^{n-1}}{\left( n-1 \right)!} & \frac{\tau^{n}}{n!} \\
         \vdots & \vdots & \vdots & \ddots & \vdots & \vdots \\
         0 & 0 & 0 & \cdots & \tau & \frac{\tau^2}{2!} \\
         0 & 0 & 0 & \cdots & 0 & \tau 
         \end{array} \right] .
    \end{split}
\end{align}
Regarding $\widetilde{\sigma}_{0,k+1}$ and $\xi_k$, they are calculated according to the following lemma:

\begin{lemma}\label{lemma2}
(\cite{carvajal2021implicit}) Let $a_l$ and $b_k$ be defined as:
\begin{align}
    \begin{split}
    a_{l}&=\frac{\tau^{n-l+1}}{\left( n-l+1 \right)!} \lambda_{l} L^{\frac{n-l+1}{n+1}},\;\; l=0,\ldots,n;\\
    b_{k}&=-\sigma_{0,k}-\sum_{l=1}^n \frac{\tau^l}{l!} z_{l,k}.  
    \end{split}
\end{align}

Then $\widetilde{\sigma}_{0,k+1}\in \mathbb{R}$ and $\xi_{k}$ is the unique pair $\left(\widetilde{\sigma}_{0,k+1},\xi_{k}\right)$ defined conforming to the following $3$ cases:
\begin{itemize}
    \item If $b_k > a_{0}$, then $\xi_{k}=\left\{ -1 \right\}$ and $\widetilde{\sigma}_{0,k+1}=-\left(r_0\right)^{n+1}\in\mathbb{R}^-$ where $r_0$ is the unique positive root of the following polynomial:
\begin{align}
    \begin{split}
        p\left( r\right)=r^{n+1}+a_{n}r^{n}+\cdots+a_1r+\left(-b_k+a_0 \right).\label{eq:pol_cas1}
    \end{split}
\end{align}

    \item If $b_k \in [-a_{0},a_{0}]$, then $\widetilde{\sigma}_{0,k+1}=0$ and $\xi_{k}=\left\{-\frac{b_k}{a_{0}}\right\}$.

    \item If $b_k < -a_{0}$, then $\xi_{k}=\left\{ 1 \right\}$ and $\widetilde{\sigma}_{0,k+1}=r_0^{n+1}\in \mathbb{R}^+$ where $r_0$ is the unique positive root of the following polynomial:
\begin{align}
    \begin{split}
        p\left( r\right)=r^{n+1}+a_{n}r^{n}+\cdots+a_1r+\left(b_k+a_0 \right).\label{eq:pol_cas2}
    \end{split}
\end{align}
\end{itemize}
\end{lemma}

Note that under the assumption of a constant sampling time $\tau$, $a_l$, and the elements of $\bm{\Phi}\left( \tau \right)$ and $\bm{B}^\ast \left( \tau \right)$ can be calculated previous to the implementation of HIDD. Opposite to the explicit discrete-time differentiators, HIDD requires estimating roots of different polynomials each iteration. They are estimated by using the Halley's method, which implies an evaluation of the polynomials \eqref{eq:pol_cas1}, \eqref{eq:pol_cas2} and its first two derivatives multiple times.  

\subsection{Halley's Method}

As it was mentioned previously, HIDD requires estimating $r_0$. In \cite{carvajal2021implicit} the Halley's method was proposed as a good alternative, which experimentally needed $3$ iteration to estimate $r_0$. It is applied as follows:
\begin{align}
    \begin{split}\label{eq:Halley}
        r_{0,j+1}=r_{0,j}-\frac{2\frac{dp(r)}{dr}|_{r=r_{0,j}}p(r_{0,j})}{2\left(\frac{dp(r)}{dr}|_{r=r_{0,j}}\right)^2-\frac{d^2p(r)}{dr^2}|_{r=r_{0,j}}p(r_{0,j})}.
    \end{split}
\end{align}

As $p\left(\left(b_{k}-a_{0}\right)^{\frac{1}{n+1}}\right)>0$ and $p\left(\left(-b_{k}-a_{0}\right)^{\frac{1}{n+1}}\right)>0$ for the polynomials \eqref{eq:pol_cas1} and \eqref{eq:pol_cas2}, then $r_{0} \in\left[0,\left(b_{k}-a_{0}\right)^{\frac{1}{n+1}}\right]$ and $r_{0} \in\left[0,\left(-b_{k}-a_{0}\right)^{\frac{1}{n+1}}\right]$, respectively. The above matches with the Cauchy's bound for the roots of a polynomial \cite{mcnamee2007numerical}. In \cite{carvajal2021implicit} was demonstrated that the Halley's method converges monotonically to $r_0$ with an convergence order $3$, \cite{mcnamee2013numerical}, for any initial $r_{0,0}$ belonging to the previous sets of initial conditions. Hence, the following initial conditions were used:

\begin{align}
    \begin{split}
        r_{0,0}&=\left( \frac{b_{k}-a_0}{2} \right)^{1 /(n+1)},\;\;\;\;\;\;\;\; \text{for} \;\; b_k>a_0, \\
        r_{0,0}&=\left( \frac{-b_k-a_0}{2}\right)^{1 /(n+1)}, \;\;\;\;\; \text{for} \;\; b_k<-a_0.
    \end{split}
\end{align}

\section{Main Results}\label{sec:3Mainresults}
As it was mentioned previously, the objective of this work is to reduce the time complexity of HIDD. First, the variables $\phi_{i}$ and $\bar{b}^{\ast}$ are defined as:

\begin{align}
    \begin{split}\label{eq:constantes_bphi}
        \phi_i&=\frac{\tau^{i-1}}{(i-1)!}, \\
        \bar{b}^{\ast}_{i,j}&=\frac{\tau^{j+1-i}}{(j+1-i)!}\lambda_{n-j+1}L^{\frac{j}{n+1}}, 
    \end{split}
\end{align}
for $i=1,2,\cdots,n+1$ and $ j=i,i+1,i+2,\cdots,n+1$.
It allows to rewrite \eqref{eq:HIDD} as follows:

\begin{itemize}
    \item If $b_k > a_{0}$,
    \begin{align}
    \begin{split}\label{eq:direct_eval_pol1}
        z_{i,k+1}&=\sum_{j=i}^{n} \phi_{j-i+1} z_{j,k}+\bar{b}^{\ast}_{i+1,j+1}r_0^{n-j},\\
        i&=0,1,\cdots,n.
    \end{split}
    \end{align}
    
    \item If $b_k \in [-a_0,a_0]$,
    \begin{align}
    \begin{split}\label{eq:direct_eval}
        z_{0,k+1}&=b_k +\sum_{j=0}^{n} \frac{\tau^{j}}{j!}z_{j,k},\\ %%\bar{b}^{\ast}_{1,n+1} \left(\frac{b_k}{a_0} \right)
        z_{i,k+1}&=\bar{b}^{\ast}_{i+1, n+1} \left(\frac{b_k}{a_0} \right)+\sum_{j=i}^{n} \phi_{j-i+1}z_{j,k},\\
        i&=1,2,\cdots,n.
    \end{split}
    \end{align}
    
    \item If $b_k < -a_{0}$,
    \begin{align}
    \begin{split}\label{eq:direct_eval_pol2}
        z_{i,k+1}&=\sum_{j=i}^{n} \phi_{j-i+1} z_{j,k}-\bar{b}^{\ast}_{i+1,j+1}r_0^{n-j},\\
        i&=0,1,\cdots,n.
    \end{split}
    \end{align}
\end{itemize}

\subsection{Direct Evaluation}

The number of additions and subtraction, $N_A$ and the number of multiplications and divisions, $N_M$, needed to evaluate $z_{i,k+1}$ directly, after obtain $r_0$, are calculated as:

\begin{align}
    \begin{split}
        &N_{A1}(n)=(n+1)^2,\\
        &N_{M1}(n)=\frac{n^3}{6}+n^2-\frac{1}{6}n-1.
    \end{split}
\end{align}
Therefore, taking into account the $(n+1)$ assignations of $z_{i,k+1}$, the time complexity is given as:

\begin{align}
    \begin{split}
        T_1(n)=\frac{n^3}{6}+2n^2+\frac{17}{6}n+1,
    \end{split}
\end{align}
which is a cubic time. On the other hand, the Halley's method is used recursively each iteration. To evaluate the derivatives and its derivatives, one could storage the following variables to reduce the number of operations.

\begin{align}
    \begin{split}\label{eq:constant_cd}
        c_{n+1}&=n+1,\\
        c_{i}&=ia_{i},\;\;\;\;\;\;\;\;\;\;\;\; \text{for}\;\; i=1,2,\cdots,n;\\
        d_{n+1}&=n(n+1),\\
        d_{i}&=i(i-1)a_{i},\;\; \text{for}\;\; i=2,3,\cdots,n.\\
    \end{split}
\end{align}

Additionally, $\bar{j}$ is defined as the number of iteration used to estimate $r_0$, then the number of additions, subtraction, multiplications and divisions used to evaluate the polynomials and its derivatives are given as: 

\begin{align}
    \begin{split}
        N_{A2}(n)&=\bar{j}\left( 3n+1\right),\\
        N_{M2}(n)&=\bar{j}\left( \frac{3}{2}n^2+\frac{3}{2}n \right).
    \end{split}
\end{align}

Since Halley's method is implemented 3 times each iteration for HIDD, $\bar{j}=3$. Therefore, taking into account the assignation of values, the evaluation of \eqref{eq:Halley}, its initialization and comparatives:

\begin{align}
    \begin{split}\label{eq:polynoEval_Direct}
         T_2(n)=\frac{9}{2}n^2+\frac{27}{2}n+46.
         %T_2(n)=\bar{j}\left( \frac{3}{2}n^2+\frac{9}{2}n+13 \right).
        %T_2(n)=\bar{j}\left( \frac{3}{2}n^2+\frac{9}{2}n+13 \right). 4 extra addition, 8 operation of evaluate r_0, %1 comparative, falta assignation of $m=1$
        %Comparativa, asignacion y suma de m para contar 3, 7
    \end{split}
 \end{align}
where one of the operations is a $(n+1)$-th root and a for-loop was considered. Hence, the complexity of the algorithm is cubic and it is defined as:

\begin{align}
    \begin{split}
        T(n)=\frac{n^3}{6}+\frac{13}{2}n^2+\frac{110}{6}n+48.
    \end{split}
\end{align}
where the multiplications and subtraction needed to evaluate $b_k$ were taking into account.

\subsection{Horner Method}
 Although the variables $\phi_{\cdot}$, $\bar{b}^{\ast}_{\cdot,\cdot}$, $c_{\cdot}$, $d_{\cdot}$ reduces the number of basic operations, it does not reduce time complexity of the realization \eqref{eq:HIDD} with respect to $n$. Based on the Horner methodology, one could calculate $z_{i,k}$ as follows:
 \begin{itemize}
   \item If $b_k>a_0$
     \begin{align}
       \begin{split}\label{eq:Horner_1_1}
        &z_{i,k+1}=\sum_{j=i}^{n} \phi_{j-i+1} z_{j,k}+\ldots\\
        &\ldots+(\cdots((\bar{b}^{\ast}_{i+1,i+1})r_0+\bar{b}^{\ast}_{i+1,i+2})\cdots)r_0+\bar{b}^{\ast}_{i+1,n+1}.\\
        &i=0,1,\cdots,n.
    \end{split}
 \end{align}
   \item If $b_k<-a_0$
     \begin{align}
       \begin{split}\label{eq:Horner_1_2}
        &z_{i,k+1}=\sum_{j=i}^{n} \phi_{j-i+1} z_{j,k}-\ldots\\
        &\ldots-(\cdots((\bar{b}^{\ast}_{i+1,i+1})r_0+\bar{b}^{\ast}_{i+1,i+2})\cdots)r_0+\bar{b}^{\ast}_{i+1,n+1}.\\
        &i=0,1,\cdots,n.
       \end{split}
     \end{align}
     
 \end{itemize}

This methodology presents the following number of basic operations:

\begin{align}
    \begin{split}
        N_{A3}(n)&=(n+1)^2,\\
        N_{M3}(n)&=n(n+1).
    \end{split}
\end{align}

As $\bm{\Phi}(\tau)$ and $\bm{B}(\tau)$ are Toeplitz matrix \cite{bai2000templates}, the time complexity of evaluate $z_{i,k+1}$ could be reduce to a linearithmic time ($n \log{n}$) using the discrete Fourier transform. This alternative will be analyzed in a future work. To evaluate the polynomials and its derivatives, $n$ is considered greater than $1$. Here two methodologies based on Horner's method are analyzed, the first one is evaluate the polynomials and derivatives as follows:

\begin{align}
    \begin{split}\label{eq:Horner_1}
        p\left( r\right)&=(\cdots((r+a_{n})r+a_{n-1})\cdots)r+a_0\pm b_k,\\
        \frac{d p(r)}{dr}&=(\cdots((c_{n+1}r+c_{n})r+c_{n-1})\cdots)r+c_1,  \\
        \frac{d^2 p(r)}{dr^2}&=(\cdots((d_{n+1}r+d_{n})r+d_{n-1})\cdots)r+d_2.
    \end{split}
\end{align}

The methodology \eqref{eq:Horner_1} use the following number of basic operations:

\begin{align}
    \begin{split}
        N_{A4}(n)&=\bar{j}\left( 3n+1\right),\\
        N_{M4}(n)&=\bar{j}\left( 3n-1 \right).
    \end{split}
\end{align}

Similar to \eqref{eq:polynoEval_Direct}, one obtains:

\begin{align}
    \begin{split}
         T_4(n)=18n+43.
    \end{split}
 \end{align}
 
 However, one could take advantage of the evaluation of $p(r)$ to evaluate $\frac{dp(r)}{dr}$ and  $\frac{d^2p(r)}{dr^2}$ with the following methodology for $n\geq2$:
 
\begin{align}
    \begin{split}
        F_{i+1}&= rF_{i}+a_{n-i-1},\\
        dF_{i+1}&= rdF_{i}+F_{i+1},\\
        ddF_{i+1}&= rddF_{i}+dF_{i+1},
    \end{split}
\end{align}
for $i=0,\cdots,n-3$, with $F_{0}=r+a_{n}$, $dF_{0}=r+F_{0}$, $ddF_{0}=r+dF_{0}$, and the value of the evaluation is given as:

\begin{align}
    \begin{split}\label{eq:Horner_2}
        F_{n-1}&= rF_{n-2}+a_{1},\\
        p(r)&= rF_{n-1}+a_{0}\pm b_k,\\
        \frac{dp(r)}{dr}&= rdF_{n-2}+F_{n-1},\\
        \frac{d^2p(r)}{dr^2}&=2ddF_{n-2}. 
    \end{split}
\end{align}

It increases the number of assignations and additions but reduces the multiplications, therefore, one obtains the following time complexity:

\begin{align}
    \begin{split}
         % N_m=3n-2  ,   Ns=3n+1
         T_5(n)=36n+55.
    \end{split}
\end{align}
 
Using the methodologies \eqref{eq:Horner_1_1}, \eqref{eq:Horner_1_2} and \eqref{eq:Horner_1} a quadratic time is obtained, which is defined as: 

\begin{align}
    \begin{split}
        T(n)=2n^2+24n+46.
        %T(n)=2n^2+4n+2
        %N_{A3}(n)&=(n+1)^2,\\
        %N_{M3}(n)&=n(n+1). (n+1) asignaciones
    \end{split}
\end{align}

If \eqref{eq:Horner_2} is used instead of \eqref{eq:Horner_1}, the quadratic time is given as:
\begin{align}
    \begin{split}
        T(n)=2n^2+42n+58.
        %T(n)=2n^2+4n+2
        %N_{A3}(n)&=(n+1)^2,\\
        %N_{M3}(n)&=n(n+1). (n+1) asignaciones
    \end{split}
\end{align}

\subsection{Shaw–Traub Algorithm}

Similar to the methodology \eqref{eq:Horner_2}, an algorithm designed to calculate the normalized derivatives, $\frac{1}{i!}\frac{d^{i}p(r)}{dr^{i}}$, was proposed in \cite{shaw1974number}. Even, the Horner method used to evaluate a polynomial is a special case of this algorithm. It allows to reduce the number of multiplication but add divisions, assignations and additions. In this work a modified algorithm is used, which relies on Shaw–Traub algorithm:

\begin{align}
    \begin{split}\label{eq:Shaw}
        &t_1=r, \;\; t_i=t_{i-1}r, \;\; \text{for}\;\; i=2,3,\cdots,n;\\
        &T_i^{-1}=a_{n-i} t_{n-i}, \;\; \text{for}\;\; i=0,1,\cdots,n-1;\\
        &T_n^{-1}=a_0\pm b_k,\\
        &T_0^0=t_n r, \;\; T_1^{1}=T_0^0,\;\; T_2^2=T_0^0;\\
        &T_i^j=T_{i-1}^{j-1}+T_{i-1}^j, \;\; \text{for}\;\; j=0,1,2;\;i=j+1,\cdots,n+1;\\
        &p(r)=T_{n+1}^0, \;\; \frac{dp(r)}{dr}=\frac{T^{1}_{n+1}}{t_1}, \;\; \frac{d^2p(r)}{dr^2}=2\frac{T^{2}_{n+1}}{t_2}. 
    \end{split} 
\end{align}

Since $T_{0}^0=T_{1}^1=T_{2}^2$, two assignations could be avoided if $T_{0}^0$ is used instead of $T_{1}^1$ and $T_{2}^2$. Then the algorithm \eqref{eq:Shaw} presents an linear time given as:

\begin{align}
    \begin{split}
        T_6(n)=30n+70. %10n+9  %2n+3, 3n+1, 5n+7 con asignaciones y 5n+5, para calcular todo por 3 y + 43
    \end{split}
\end{align}

Applying the algorithms \eqref{eq:Horner_1_1}, \eqref{eq:Horner_1_2} and \eqref{eq:Shaw}, HIDD presents a quadratic time, which is given as:

\begin{align}
    \begin{split}\label{eq:time_Shaw}
        T(n)=2n^2+36n+74.  %2n^2+3n+1  la suma mas n+2
    \end{split}
\end{align}

It is important to mention that in \cite{shaw1975analysis}, the best parameters for the family of algorithms presented in \cite{shaw1974number} were founded. Then the time complexity \eqref{eq:time_Shaw} could be reduced tuning its parameters but not its order. As Halley's method does not use higher-order derivatives, this work will not consider the algorithm presented in \cite{de1975improved}, which improve the algorithm proposed in \cite{shaw1974number} for the $n+1$ normalized derivatives.  

\section{Simulation}\label{sec:4Simulations}

In this section, two comparatives are presented. The first one correspond to a graphical comparative of the required basic operations of the algorithms. The second one is a comparative of the time of simulation for the 4 methodologies proposed in this work. The methodology composed of \eqref{eq:Halley}, \eqref{eq:direct_eval_pol1}, \eqref{eq:direct_eval}, \eqref{eq:direct_eval_pol2} and a direct evaluation of the polynomials is referenced as direct evaluation, the methodology composed of \eqref{eq:Halley}, \eqref{eq:direct_eval} \eqref{eq:Horner_1_1}, \eqref{eq:Horner_1_2} and \eqref{eq:Horner_1} is referenced as half-Horner algorithm, the methodology composed of \eqref{eq:Halley}, \eqref{eq:direct_eval},  \eqref{eq:Horner_1_1}, \eqref{eq:Horner_1_2} and \eqref{eq:Horner_2} is referenced as Full-Horner algorithm and \eqref{eq:Halley}, \eqref{eq:direct_eval}, \eqref{eq:Horner_1_1}, \eqref{eq:Horner_1_2} and \eqref{eq:Shaw} is referenced as Shaw-Traub algorithm.

\subsection{Simulation I}

\begin{figure}
    \centering
    \includegraphics[width=.95\linewidth]{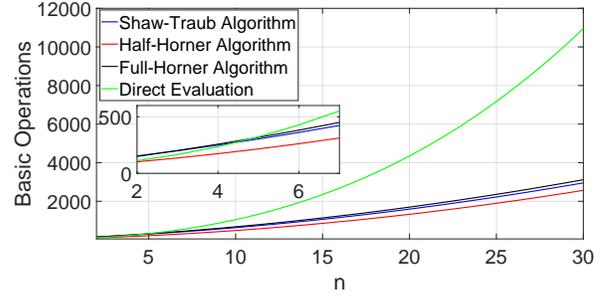}
    \caption{Comparative between the time complexity of the methodologies.}
    \label{fig:SimulationI}
\end{figure}

In this simulation the time complexity of the methodologies are evaluated for $2\leq n\leq 30$. The results are presented in Figure \ref{fig:SimulationI}. One can see that the algorithm with less basic operations is Half-Horner algorithm. Furthermore Shaw-Traub and Full-Horner algorithm have less basic operations than direct evaluation for $n>4$. It is important note the difference of Direct evaluation with respect to the other algorithms for a value of $n>=7$, $240$ or more basic operations with respect to Half-Horner algorithm.

\subsection{Simulation II}
Since the methodologies have a different proportion of basic operations, this simulation aims to compare the time require to simulate the algorithms. Additionally, an simulation without the parameters defined in Equations \eqref{eq:constantes_bphi} and \eqref{eq:constant_cd} is simulated, which is the same methodology than the direct evaluation without the parameters $\phi_i$, $\bar{b}^{\ast}_{i,j}$, $c_i$ and $d_i$. Here $n=3$, $n=7$ and $n=10$ are considered, the values of the signal with noise and the constants defined in \eqref{eq:constantes_bphi} and \eqref{eq:constant_cd} are calculated previous to the simulation.  The sampling time is selected as $\tau=0.001\sec$ and $t=2000,\;10000,\;25000,\;50000\sec$. The results are presented in Tables \ref{tab:numericalresults1}-\ref{tab:numericalresults3}. The most efficient methods with respect to the simulation time were the Half-Horner and Full-Horner algorithms, both present a similar performance for $n=3$, $n=7$ and $n=10$. For $n=3$, direct evaluation has a better performance than Shaw-Traub, it contrasts to the results obtained for $n=10$, where Shaw-Traub algorithm reduces the simulation time compared to direct evaluation. The above fact matches with its time complexity.

\begin{remark}
Half-Horner and Full-Horner algorithms reduced the simulation time more than 25 times for $n=10$. It can be seen in Table \ref{tab:numericalresults3}. It comes from the use of the variables $\bar{b}^{\ast}_{i,j}$, $\phi_{i}$, $c_i$ and $d_i$ and a reduction of the time complexity. 
\end{remark}

\begin{table}[H]
\centering
\begin{tabular}{|l|l|l|l|l|l}
\cline{1-5}
\textbf{} & $2000\sec$ & $10000\sec$ & $25000\sec$ & $50000\sec$ & \\ \cline{1-5}
\textbf{Evaluation} &   & & &  &  \\ 
\textbf{without} $\phi_i$, & $0.5963\sec$  & $2.990\sec$ & $7.470\sec$ &$15.059\sec$&  \\ 
$\bar{b}^{\ast}_{i,j}$,\;$c_i$ and $d_i$ &   & & &  &  \\  \cline{1-5}
\textbf{Direct}  &   & & &  &  \\ 
\textbf{Evaluation}. & $0.3578\sec$  & $1.783\sec$ & $4.475\sec$ & $9.027\sec$&  \\ \cline{1-5}
\textbf{Half-Horner}. & $0.3494\sec$  & $1.753\sec$ & $4.411\sec$ & $8.896\sec$& \\ \cline{1-5}
\textbf{Full-Horner}. & $0.3496\sec$ & $1.756\sec$ & $4.407\sec$ &$8.908\sec$&  \\ \cline{1-5}
\textbf{Shaw-Traub}. & $0.3852\sec$ & $1.922\sec$ & $4.813\sec$ &$9.661\sec$&  \\ \cline{1-5}
\end{tabular}
\caption{Simulation time of the algorithms for $n=3$ and $\tau=0.001\sec$.}
\label{tab:numericalresults1}
\end{table}

\begin{table}[H]
\centering
\begin{tabular}{|l|l|l|l|l|l}
\cline{1-5}
\textbf{} & $2000\sec$ & $10000\sec$ & $25000\sec$ & $50000\sec$ & \\ \cline{1-5}
\textbf{Evaluation} &   & & &  &  \\ 
\textbf{without} $\phi_i$, & $6.791\sec$  & $33.808\sec$ & $85.045\sec$ & $169.95\sec$ &  \\ 
$\bar{b}^{\ast}_{i,j}$,\;$c_i$ and $d_i$ &   & & &  &  \\  \cline{1-5}
\textbf{Direct}  &   & & &  &  \\ 
\textbf{Evaluation}. & $0.486\sec$  & $2.414\sec$ & $6.035\sec$ & $12.51\sec$ &  \\ \cline{1-5}
\textbf{Half-Horner}. & $0.466\sec$  & $2.286\sec$ & $5.729\sec$ & $11.61\sec$ & \\ \cline{1-5}
\textbf{Full-Horner}. & $0.457\sec$ & $2.293\sec$ & $5.763\sec$ & $11.51\sec$ &  \\ \cline{1-5}
\textbf{Shaw-Traub}. & $0.503\sec$ & $2.46\sec$ & $6.210\sec$ & $12.718\sec$ &  \\ \cline{1-5}
\end{tabular}
\caption{Simulation time of the algorithms for $n=7$ and $\tau=0.001\sec$.}
\label{tab:numericalresults2}
\end{table}

\begin{table}[H]
\centering
\begin{tabular}{|l|l|l|l|l|l}
\cline{1-5}
\textbf{} & $2000\sec$ & $10000\sec$ & $25000\sec$ & $50000\sec$ & \\ \cline{1-5}
\textbf{Evaluation} &   & & &  &  \\ 
\textbf{without} $\phi_i$, & $14.312\sec$  & $71.6\sec$ & $179.37\sec$ & $358.09\sec$ &  \\ 
$\bar{b}^{\ast}_{i,j}$,\;$c_i$ and $d_i$ &   & & &  &  \\  \cline{1-5}
\textbf{Direct}  &   & & &  &  \\ 
\textbf{Evaluation}. & $0.831\sec$  & $4.192 \sec$ & $10.33\sec$ & $20.527\sec$ &  \\ \cline{1-5}
\textbf{Half-Horner}. & $0.5437\sec$  & $2.75\sec$ & $6.858\sec$ & $13.692\sec$ & \\ \cline{1-5}
\textbf{Full-Horner}. & $0.5631\sec$ & $2.807\sec$ & $6.997\sec$ & $14.139\sec$ &  \\ \cline{1-5}
\textbf{Shaw-Traub}. & $0.6101\sec$ & $3.097\sec$ & $7.767\sec$ & $15.464\sec$ &  \\ \cline{1-5}
\end{tabular}
\caption{Simulation time of the algorithms for $n=10$ and $\tau=0.001\sec$.}
\label{tab:numericalresults3}
\end{table}

\section{Conclusion}\label{sec:5Conclusion}

In this work four methodologies were designed to implement the implicit discrete-time differentiator HIDD, which rely on the Horner's method and the Shaw-Traub algorithm. 3 methodologies allow to reduce the time complexity of the implicit differentiator with respect to $n$, i.e., they present a quadratic time instead of a cubic time. The methodologies show a noticeable performance compared to an direct implementation without the parameters $\phi_i$, $\bar{b}_{i,j}$, $c_i$ and $d_i$. However, the simulations show that Half-Horner and Full-Horner algorithms reduce the number of basic operations and its simulation time. Even, they reduced the simulation time of the differentiator more than 25 times for $n=10$. As future work, the discrete Fourier transform and the methodology proposed in \cite{de1975improved} will be considered to reduce the time complexity.

%Here two methodologies based on Horner's method are analyzed, the first one correspond to apply 

\bibliographystyle{./bibliography/IEEEtran}
\bibliography{./bibliography/IEEEabrv,./bibliography/IEEEexample}

\begin{thebibliography}{10}
\providecommand{\url}[1]{#1}
\csname url@samestyle\endcsname
\providecommand{\newblock}{\relax}
\providecommand{\bibinfo}[2]{#2}
\providecommand{\BIBentrySTDinterwordspacing}{\spaceskip=0pt\relax}
\providecommand{\BIBentryALTinterwordstretchfactor}{4}
\providecommand{\BIBentryALTinterwordspacing}{\spaceskip=\fontdimen2\font plus
\BIBentryALTinterwordstretchfactor\fontdimen3\font minus
  \fontdimen4\font\relax}
\providecommand{\BIBforeignlanguage}[2]{{%
\expandafter\ifx\csname l@#1\endcsname\relax
\typeout{** WARNING: IEEEtran.bst: No hyphenation pattern has been}%
\typeout{** loaded for the language `#1'. Using the pattern for}%
\typeout{** the default language instead.}%
\else
\language=\csname l@#1\endcsname
\fi
#2}}
\providecommand{\BIBdecl}{\relax}
\BIBdecl

\bibitem{kaveh2008blood}
P.~Kaveh and Y.~B. Shtessel, ``Blood glucose regulation using higher-order
  sliding mode control,'' \emph{International Journal of Robust and Nonlinear
  Control}, vol.~18, no. 4‐5, pp. 557--569, 2008.

\bibitem{shtessel2007smooth}
Y.~B. Shtessel, I.~A. Shkolnikov, and A.~Levant, ``Smooth second-order sliding
  modes: Missile guidance application,'' \emph{Automatica}, vol.~43, no.~8, pp.
  1470--1476, 2007.

\bibitem{iqbal2010robust}
M.~{Iqbal}, A.~I. {Bhatti}, S.~I. {Ayubi}, and Q.~{Khan}, ``Robust parameter
  estimation of nonlinear systems using sliding-mode differentiator observer,''
  \emph{IEEE Transactions on Industrial Electronics}, vol.~58, no.~2, pp.
  680--689, Feb 2011.

\bibitem{Levant_HSMD}
A.~Levant, ``Higher-order sliding modes, differentiation and output-feedback
  control,'' \emph{International Journal of Control}, vol.~76, no. 9-10, pp.
  924--941, 2003.

\bibitem{Utkin2009}
V.~I. Utkin, J.~Guldner, and J.~Shi, \emph{Sliding Mode Control in
  Electro-Mechanical Systems}, 2nd~ed.\hskip 1em plus 0.5em minus 0.4em\relax
  CRC Press, 2009.

\bibitem{Miki2014}
M.~Livne and A.~Levant, ``Proper discretization of homogeneous
  differentiators,'' \emph{Automatica}, vol.~50, no.~8, pp. 2007--2014, 2014.

\bibitem{Stefan_dif}
S.~{Koch} and M.~{Reichhartinger}, ``Discrete-time equivalent homogeneous
  differentiators,'' in \emph{2018 15th International Workshop on Variable
  Structure Systems (VSS)}, July 2018, pp. {354--359}.

\bibitem{koch2019discrete}
S.~{Koch}, M.~{Reichhartinger}, M.~{Horn}, and L.~{Fridman}, ``Discrete-time
  implementation of homogeneous differentiators,'' \emph{IEEE Transactions on
  Automatic Control}, vol.~65, no.~2, pp. 757--762, Feb 2020.

\bibitem{barbot2020discrete}
J.-P. Barbot, A.~Levant, M.~Livne, and D.~Lunz, ``Discrete differentiators
  based on sliding modes,'' \emph{Automatica}, vol. 112, p. 108633, 2020.

\bibitem{carvajal2021implicit}
\BIBentryALTinterwordspacing
J.~E. Carvajal-Rubio, J.~D. S\'anchez-Torres, M.~Defoort, M.~Djemai, and A.~G.
  Loukianov, ``Implicit and explicit discrete-time realizations of homogeneous
  differentiators,'' \emph{International Journal of Robust and Nonlinear
  Control}, 2021, {Special Issue on Homogeneous Sliding-Mode Control and
  Observation}. [Online]. Available: \url{https://doi.org/10.1002/rnc.5505}
\BIBentrySTDinterwordspacing

\bibitem{mcnamee2013numerical}
J.~M. McNamee and V.~Pan, \emph{Numerical Methods for Roots of Polynomials -
  Part II}, ser. Studies in Computational Mathematics.\hskip 1em plus 0.5em
  minus 0.4em\relax Amsterdam London: Elsevier Science, 2013, vol.~16.

\bibitem{sipser2012introduction}
M.~Sipser, \emph{Introduction to the Theory of Computation}.\hskip 1em plus
  0.5em minus 0.4em\relax Cengage learning, 2012.

\bibitem{mcnamee2007numerical}
J.~M. McNamee, \emph{Numerical Methods for Roots of Polynomials - Part I}, ser.
  Studies in Computational Mathematics.\hskip 1em plus 0.5em minus 0.4em\relax
  Amsterdam London: Elsevier Science, 2007, vol.~14.

\bibitem{shaw1974number}
M.~Shaw and J.~F. Traub, ``On the number of multiplications for the evaluation
  of a polynomial and some of its derivatives,'' \emph{Journal of the ACM
  (JACM)}, vol.~21, no.~1, pp. 161--167, 1974.

\bibitem{de1975improved}
L.~De~Jong and J.~Van~Leeuwen, ``An improved bound on the number of
  multiplications and divisions necessary to evaluate a polynomial and all its
  derivatives,'' \emph{ACM SIGACT News}, vol.~7, no.~3, pp. 32--34, 1975.

\bibitem{knuth2014art}
D.~E. Knuth, \emph{Art of computer programming, volume 2: Seminumerical
  algorithms}.\hskip 1em plus 0.5em minus 0.4em\relax Addison-Wesley
  Professional, 2014.

\bibitem{golub2013matrix}
G.~H. Golub and C.~F. Van~Loan, \emph{Matrix computations}, 4th~ed.\hskip 1em
  plus 0.5em minus 0.4em\relax The Johns Hopkins University Press, Baltimore,
  2013.

\bibitem{carvajal2019discretization}
J.~E. {Carvajal-Rubio}, A.~G. {Loukianov}, J.~D. {Sánchez-Torres}, and
  M.~{Defoort}, ``On the discretization of a class of homogeneous
  differentiators,'' in \emph{2019 16th International Conference on Electrical
  Engineering, Computing Science and Automatic Control (CCE)}, September 2019,
  pp. {1--6}.

\bibitem{bai2000templates}
Z.~Bai, J.~Demmel, J.~Dongarra, A.~Ruhe, and H.~van~der Vorst, \emph{Templates
  for the solution of algebraic eigenvalue problems: a practical guide}.\hskip
  1em plus 0.5em minus 0.4em\relax Society for Industrial and Applied
  Mathematics (SIAM), 2000.

\bibitem{shaw1975analysis}
M.~Shaw and J.~F. Traub, ``Analysis of a family of algorithms for the
  evaluation of a polynomial and some of its derivatives,'' \emph{Department of
  Computer Science Carnegie Mellon University}, 1975.

\end{thebibliography}
 
 %V. Y. Pan, USSR Computational Math. and Math. Physics 2 (1963), 137–146. V. Y. Pan [Problemy Kibernetiki 5 (1961), 17–29].
 % Aunque los metodos basados en Adaptation of coefficients reducen el numero de multiplicaciones, incrementado posiblemente las sumas e asignaciones, requiere conocer los parametros a_0-b_k antes de su implementacion. Es importante mencionar que la multiplicacion podria ser reducida a 
 
\end{document}